\documentclass[twocolumn,pra]{revtex4-1}

\usepackage{graphicx}
\usepackage{amsmath}
\usepackage{amsfonts}
\usepackage{braket}

\begin{document}
	
	\title{ A dynamical interpretation of the Pauli Twirling Approximation and Quantum Error Correction }
	\author{Amara Katabarwa}
	\email{akataba@uga.edu}
	\affiliation{Department of Physics and Astronomy, University of Georgia, Athens, Georgia 30602, USA}
	
	\begin{abstract}
		 One of simplest and most widely used error model in working with quantum circuits is the Pauli Twirling Approximation (PTA). Restricting ourselves to analysis of free dynamics of qubits we show explicitly how application of PTA is equivalent to ignoring most of the quantum back action of the system and give a general argument as to why this approximation leads to low logical error rates in fault tolerant stabilizer circuits as compared to other quantum channels. We provide numerical evidence that PTA's performance in modeling noise gets worse as number of qubits increase.
	\end{abstract}
 \maketitle
 \section{Introduction}
 It is known that simulation of quantum circuits with noise is an exponentially difficult task on a classical computer. This hugely restricts the size of the system that can be studied and presents a considerable stumbling block to calculating accurate quantities like logical error rates or error thresholds. By the Gottesman-Knill theorem \cite{Gottesmannarxiv} restriction of the gates in the circuit to elements from the Pauli group and its normalizer while only measuring operators from Pauli group ensures the efficient classical simulability of the quantum circuit. This has to do with the isomorphism that can be realized between elements of the Pauli  group with matrix multiplication being the group operation and a vector space \(\mathbb{F}_2^n \) (vector space over the field \(\mathbb{Z}_2 \)) with a symplectic product\cite{Calderbank1997,Gaitan2008,Gottesmanthesis}. The sacrifice that one makes for the restriction is that one is restricted to modeling errors by elements for the Pauli Group which certainly will not reflect the true noisy behavior. \\ 
 \indent A natural question then arises, namely how good is this model and what other simple models can be used to approximate the true dynamics of the noise. First steps were taken by  Magesan \textit{et al.}  and Guti\'errez \textit{et al.}\cite{Mauricio2013,Easwar2013} providing noise channels with the desire that they provided a least upper bound to logical error rates  with respect to some measure. These dealt with the case of one qubit but later work \cite{DaneilPuzzuoli2014,GellerZhou2013,Tomita2014,Gutierrez2015,Katabarwa2015}  extended the study to more than one qubit. For this work, we take a step back and consider the simplest of the error models on offer so far and that is the Pauli Twirling Approximation (PTA). 
 
  \section{Pauli Twirling Approximation and Lindblad Forms}
  The time evolution of a density matrix \( \rho \) represented by some superoperator \( \Lambda \) is known to be given by:
  \begin{equation} \label{eq:1}
  \rho \rightarrow \Lambda(\rho) = \sum\limits_{m} E_m \rho E_m^{\dagger},
  \end{equation} 
  where the \(E_m\) are \(N \times N\) Kraus matrices. Next consider a finite set of operations \( \mathcal{B}=\{B_m\} \) from some representation of a group ,with $m = 1, \dots, K.$ \textit{Twirling} \cite{Emerson2007,Bendersky2008,Lopez2009,Lopez2010} the channel to obtain a new channel \( \tilde{\Lambda} \) is to perform the following:
  \begin{equation} \label{eq:2}
  \tilde{\Lambda} = \frac{1}{K} \sum\limits_{m=1}^{m=K} B_m^\dagger \Lambda(B_m \rho B_m^\dagger) B_m.
  \end{equation}
  In order to derive PTA, we assert the set \( \mathcal{B}\) to be the n-qubit Pauli basis \( \mathcal{P}_n\), defined as consisting of all possible tensor products and the write the Kraus matrices in terms of this basis
  
  \begin{equation} \label{eq:3}
  \mathcal{P}_n = \{I,X,Y,Z\}^{\otimes^n},
  \end{equation}
  giving a total of \(4^n\) distinct elements. Performing the twirl gives \( \tilde{\Lambda}\) that is diagonal in the Pauli basis, namely
  \begin{equation} \label{eq:4}
  \tilde{\Lambda} = \sum\limits_{B_m \in \mathcal{P}_n} p_m B_m \rho B_m.
  \end{equation}
  
  If the kraus \(E_m\) matrices are written in the Pauli basis we see from 
  comparing (\ref{eq:1}) with (\ref{eq:4}) that we are simpling ignoring the \textit{cross terms} i.e the non-PTA terms. Concretely, we introduce a model which shall be used throughout the rest of this work. Consider the simple case of a single qubit undergoing amplitude damping with relaxation time \(T_1\). The kraus matrices are 
  \begin{align}
  \label{eq:5}
  \begin{split}
  E_1 & = \begin{pmatrix}
  1  &  0  \\
  0  &  \sqrt{1- \lambda}
  \end{pmatrix},
  \\
  E_2 & = \begin{pmatrix}
  0  &  \sqrt{\lambda}  \\
  0  &  0
  \end{pmatrix},
  \end{split}
  \end{align}
  with \(\lambda = 1 - e^{-\frac{t_{step}}{T_1}} \). Then for this simple model (\ref{eq:1}) becomes
  \begin{equation} \label{eq:6}
  \begin{split}
  \Lambda(\rho)= \frac{\lambda}{4} X \rho X - \frac{i \lambda}{4}X \rho Y + \frac{i \lambda}{4} Y \rho X + \frac{\lambda}{4}Y \rho Y +  \\ \frac{2+ 2\sqrt{1-\lambda}-\lambda}{4} \rho + \frac{\lambda}{4} \rho Z + \frac{2 - 2\sqrt{1-\lambda}-\lambda}{4} Z\rho Z .
  \end{split}
  \end{equation}
  Performing the twirling process removes the cross terms to give 
  \begin{equation}
  \begin{split}
  \Lambda(\rho)=  \frac{2+ 2\sqrt{1-\lambda}-\lambda}{4} \rho + \frac{\lambda}{4} X \rho X + \frac{\lambda}{4}Y \rho Y + \\ \frac{2 - 2\sqrt{1-\lambda}-\lambda}{4} Z\rho Z 
  \end{split}
  \end{equation}
  \indent Often in physics when we make an approximation we have a physical picture or justification in mind e.g the system is in the weak coupling limit with respect to some parameter thus justifying a perturbative expansion or a special configuration of the system allows us to ignore terms in Hamiltonians. The goal for this work is to understand when it is ok to jump from (6) to (7) for free dynamics of qubits. Of course the ultimate goal is to understand why the jump from (6) to (7) is justified within the context of stabilizer circuits, which is the main thrust of this work. \\

 \subsection{From Kraus to Lindblad}
 \indent To get a better sense of what we do when we throw away the cross terms, the key step is to transform the Kraus representation in  (\ref{eq:1}) to a form of the Lindblad master equation. 
 In this section we review the method introduced by Andersson \textit{et. al} \cite{erikaanderssonjamesd.cressermichaelj.w.hall2007} that accomplishes the production of a Lindblad master equation from a Kraus representation and vice versa. Let \(\phi  \) be a completely positive map and \(\rho\) a density matrix. We can represent the density matrix under the action of the map and the original density matrix in terms of a Hermitian orthonormal basis i.e 
 \begin{align}
 G_a^{\dagger} = G_a , \\
 tr(G_a G_b) = \delta_{ab}.
 \end{align}
 In this basis we have 
 \begin{align}
 \phi(\rho) &= \sum_{k} tr \left(\phi(\sum_{l}tr\{G_l\rho\}G_l) \right) G_k ,\\
 &= \sum_{kl} F_{kl}r_{l} G_k,
 \end{align}
 
 where \( F_{kl}= tr\left( G_k tr(\phi(G_l)) \right)  \) and \(r_l =tr(G_l \rho) \)
 
 The \(\phi \) induces time evolution i.e \( \rho(t) = \phi[\rho(0)] \) so that if we take the time derivative we have \( \dot{\rho(t)} = \dot{\phi}[\rho(0)] \). Therefore we have the following equation
 
 \begin{equation}
 \dot{\rho(t)} = \sum_{kl} \dot{F_{kl}}r_{l}(0) G_k.
 \end{equation}
 
 We now suppose that the evolution of a density matrix \(\rho \) is also governed by a master equation, 
 \begin{equation}
 \dot{\rho} = \Lambda(\rho),
 \end{equation}
 this naturally leads to analogue of the  \(F\) matrix denoted as \(L\) defined in the following manner
 
 \begin{equation}
 L_{kl} = tr(G_k\Lambda(G_l) ),
 \end{equation}
 
 This leads to a similar summation as (12) but this time we have that
 \begin{equation}
 \dot{\rho} = \sum_{klm} = L_{kl}F_{ml}r_m(0) G_k .
 \end{equation}
 
 Comparing (12) and (15) we arrive at the result
 \begin{equation}
 \dot{F} = LF ,
 \end{equation}
 
 or to put into terms of the super-operators we have
 \begin{equation}
 \dot{\phi} = \Lambda \circ \phi ,
 \end{equation}
 or 
 \begin{equation}
 \dot{\phi} \circ \phi^{-1} = \Lambda .
 \end{equation}
 
 Using the Jamiolkowski isomorphism we can represent the map  \(\Lambda \) in terms of its choi matrix \(R_{ef}\)\cite{Arrighi2004} which in turn is calculated in terms of the kraus evolution map i.e 
 \begin{equation}
 \dot{\rho} = \sum_{ef} R_{ef} \bra{e_2} \rho \ket{f_2} \ket{e_1} \bra{f_1},
 \end{equation}
 where 
 \begin{equation}
 R_{ef} = \bra{e_1} \dot{\phi} \circ \phi^{-1}\left(\ket{e_2}\bra{f_2} \right) \ket{f_1},
 \end{equation}
 
 and indices \(e,f\) represent the following ordered pair indices \( e =\{ e_1,e_2\} \hspace{5mm} f =\{f_1,f_2\} \) while \(\{ \ket{e_1}\}, \{\ket{f_1} \}, \{ \ket{e_2}\},\{ \ket{f_2} \} \) all form an orthonormal basis for an n-dimensional Hilbert space. We can form an orthonormal non-Hermitian basis for operators on the Hilbert space as follows \( \tau_e = \ket{e_1}\bra{e_2} \). Then (19) can then be written as 
 \begin{equation}
 \dot{\rho} = \sum_{ef} R_{ef} \tau_{e} \rho \tau^{\dagger}_{f}.
 \end{equation}
 
 Lastly, we can find a representation of the choi matrix in terms of the hermitian and non-hermitian matrices of the operators on the Hilbert space. First note that 
 \begin{equation}
 \phi(\ket{e_2}\bra{f_2}) = \sum_{bc} \bra{f_2}G_c \ket{e_2} F_{cb} G_b,
 \end{equation}
 which is got expanding the non-hermitian basis in terms of Hermitian basis and applying the map \( \phi \).
 So we have
 \begin{align}
 \dot{\phi} \circ \phi^{-1}\left(\ket{e_2}\bra{f_2} \right) &= \sum_{bc} \bra{f_2}G_c\ket{e_2}\tilde{F}_{cb}\dot{\phi}(G_b),  \\
 & = \sum_{bcd}  \bra{f_2}G_c\ket{e_2}\dot{F}_{bd}\tilde{F}_{cb} G_d .
 \end{align}
 therefore
 \begin{align}
 R_{ef} = \sum_{cd}\left( \dot{F}\tilde{F}\right)_{cd} tr\left[\tau_f^{\dagger}G_c\tau_e G_d \right]  .
 \end{align}
 Consequently taking the \(\tau_a \) basis to be \(\{\ket{0}\bra{0},\ket{1}\bra{1},\ket{0}\bra{1},\ket{1}\bra{0} \} \) and hermitian basis to be the Pauli basis , carrying out calculations for kraus matrices in (5)  gives
 \begin{align}
 \dot{\rho}=  -\frac{\dot{f}}{f(t)}\left( 2 \sigma_{-} \rho \sigma_{+} -  \{\sigma_{+}\sigma_{-},\rho \} \right) , 
 \end{align}
 where \(f(t)= \sqrt{1-\lambda}\).  \\
 
 We  now focus our efforts on understanding the consequences of the term, \(\frac{\dot{f}}{f(t)}\sigma_{+}\sigma_{-} \). This is the term that appears in the anti-commutator with the density matrix and precisely a term like this is what we lack when do PTA. In fact it can be shown that the master equation(written in pauli basis) corresponding PTA map is 
 
 \begin{equation}
 \dot{\rho} = \gamma_1 \rho + \gamma_2 \sigma_x \rho \sigma_x + \gamma_3 \sigma_y \rho \sigma_y + \gamma_4 \sigma_z \rho \sigma_z .
 \end{equation}
 
 \subsection{Anti-commutator term}
 
 Since this is term that is absent when we make the Pauli Twirling Approximation we can investigate its effect on the dynamics. To do so we assume we have a single qubit in excited state freely evolving with the Hamiltonian \(H_{free} = - \sigma_z \). We are interested in the following quantity
 \begin{equation}
 p = tr(e^{-iH_{\mathbb{C}}}\rho e^{iH_{\mathbb{C}}^{\dagger}})
 \end{equation}
 where \( H_{\mathbb{C}} = -\sigma_z  -\frac{i\gamma}{2}\sigma_{+}\sigma_{-}   \) and \( \gamma=\frac{\dot{f}}{f(t)}\). Note that this effective Hamiltonian is not Hermitian and turns out to be trace decreasing.
 
 Why might one be interested in this quantity? It is the probability  of observing no excitations or putting it differently it is "measure" of the quantum backaction. PTA does not model the quantum-back action of the system properly and moreover gets rid of the Pauli error correlations present in the amplitude damping Kraus representation. Within in the stabilizer formalism, the assumed error model is that we simply have Pauli errors. We might then ask how long can we get away with assuming Pauli errors. In other words is there a period of time the error model in stabilizer formalism  gives evolution that does not differ that much from the exact noise?
  \onecolumngrid
 \begin{center}
 	\begin{figure}[h]
 		\centering
 		\includegraphics[width=5in]{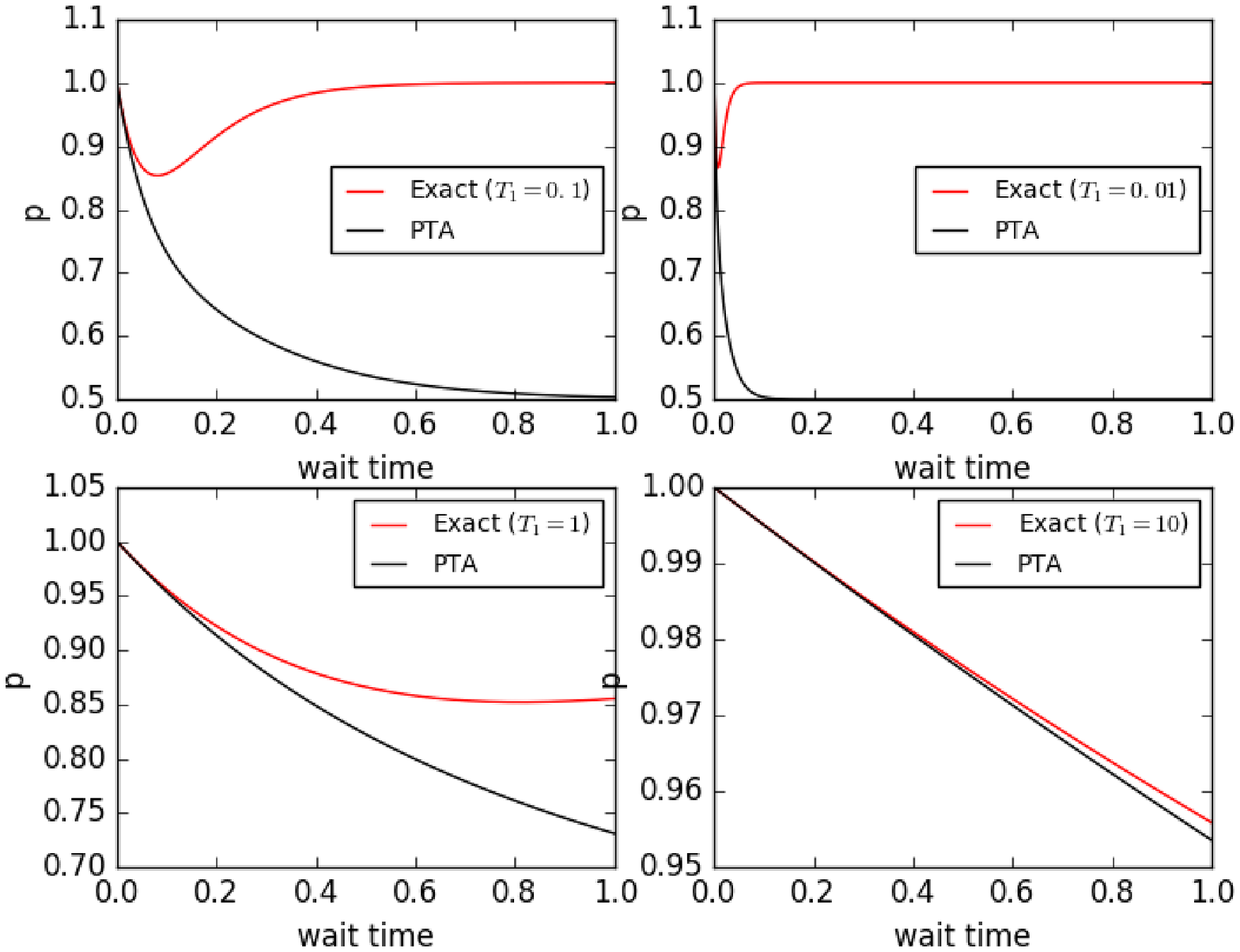}
 		\caption{Probability of seeing no excitation as function of waiting time}
 	\end{figure}
 \end{center}
 \twocolumngrid
  From the first figure, we see that for small times  the PTA model for the quantum back-action predicts approximately the same probability of no excitations as the exact amplitude damping map. Observe that we are using the wait time or integrated time and as such the probability of no excitation begins from 1. If one does not look at the system at all, the probability of observing no excitation is surely 1. Since the qubit starts in the excited state, the probability of observing no excitation decreases. But as the probability of being in the ground state increases the probability of seeing no excitation increases again.  This turn around point occurs roughly at \(t \approx T_1 \). \newline \indent
   We can also understand the long time behavior of PTA in a simple manner. Merely compose the PTA map with itself \(n\) times  and ask what it looks like i.e what is \(p_i^{(n)}\) in the following equation
   \begin{align}
   \Lambda \circ \Lambda \dots \circ \Lambda(\rho) =  p_1^{(n)} \rho + p_2^{(n)}X \rho X +p_3^{(n)}Y \rho Y +p_4^{(n)}Z \rho Z
   \end{align}
   
   The equations for \(p_i^{(n)} \) can easily be found to be the following:
   
   \begin{align}
   p_1^{(n)} &= p_1^{(1)}p_1^{(n-1)} + p_2^{(1)}p_2^{(n-1)}+p_3^{(1)}p_3^{(n-1)} +p_4^{(1)}p_4^{(n-1)} \\
   p_2^{(n)} &= p_1^{(n-1)}p_2^{(1)} + p_2^{(n-1)}p_1^{(1)}+p_3^{(n-1)}p_4^{(1)} +p_4^{(n-1)}p_3^{(1)} \\
   p_3^{(n)} &= p_1^{(n-1)}p_3^{(1)} + p_3^{(n-1)}p_1^{(1)}+p_4^{(n-1)}p_2^{(1)} +p_2^{(n-1)}p_4^{(1)} \\
   p_4^{(n)} &= p_1^{(n-1)}p_4^{(1)} + p_4^{(n-1)}p_1^{(1)}+p_3^{(n-1)}p_2^{(1)} +p_2^{(n-1)}p_3^{(1)}
   \end{align}
   Now the long time behavior which corresponds to high values of \(n\) is a fixed point.Numerically solving these equations gives that \(p_i^{(n)} = 0.25\) for high values of \(n\). The probability of observing no excitations is just the first and last terms in (29) which is gives a combined probability of 0.5. We also see one more important feature namely the PTA gets worse earlier with higher decoherence rates.
  
   Next, we could make a preliminary study of how the back action behaves as the number of qubits increases. For this discussion we pick an admittedly crude and arbitrary measure of this, namely we simply ask at what first point in time does the difference in probabilities predicted differ by more than the probability of a bit flip. 
   \begin{figure}
   	   \includegraphics[width=4in]{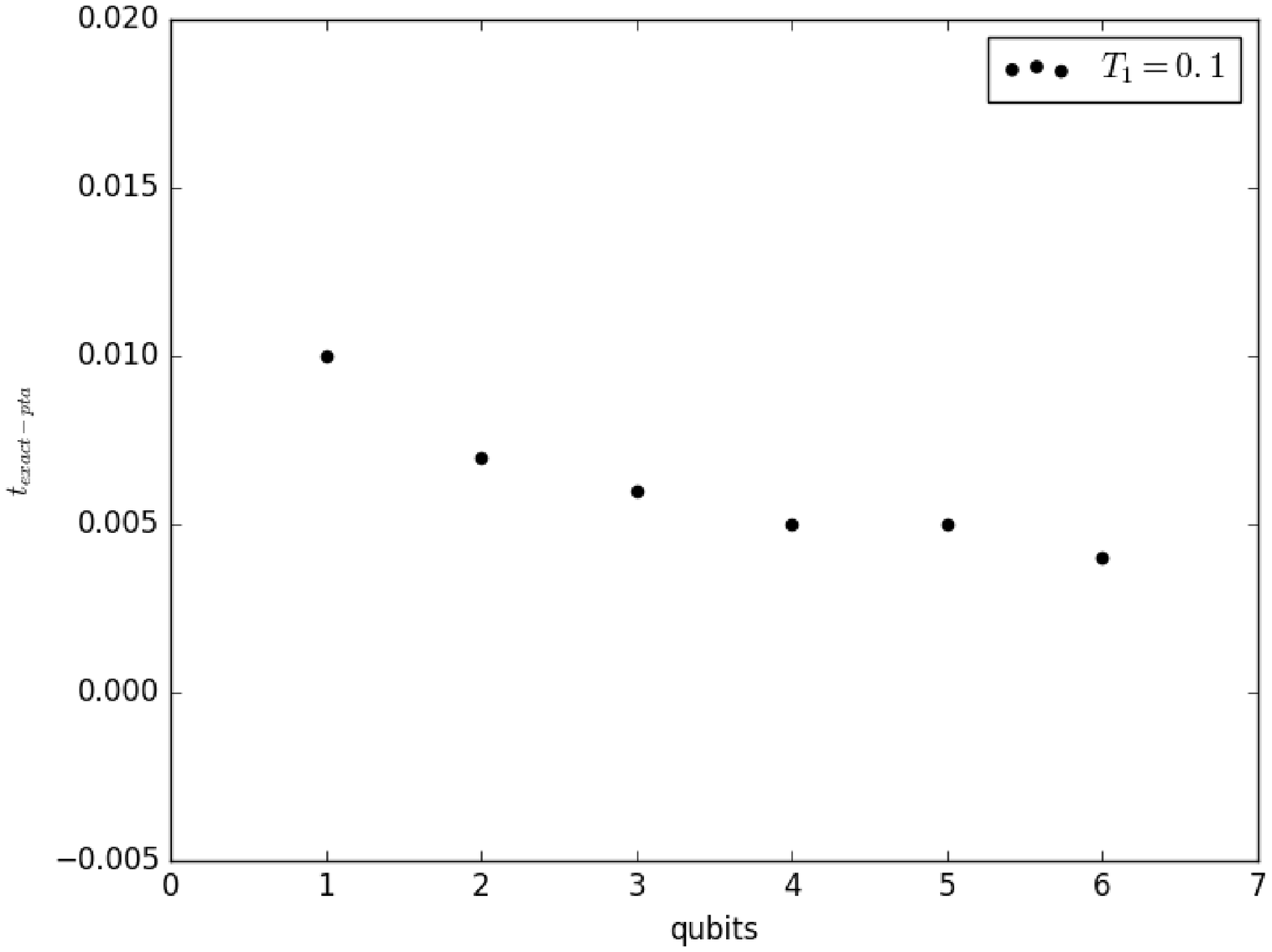}
      \caption{First time at which probabilities differ by more than probabilities of a bit flip}
   \end{figure}
   
   \begin{figure}
   	  \includegraphics[width=4in]{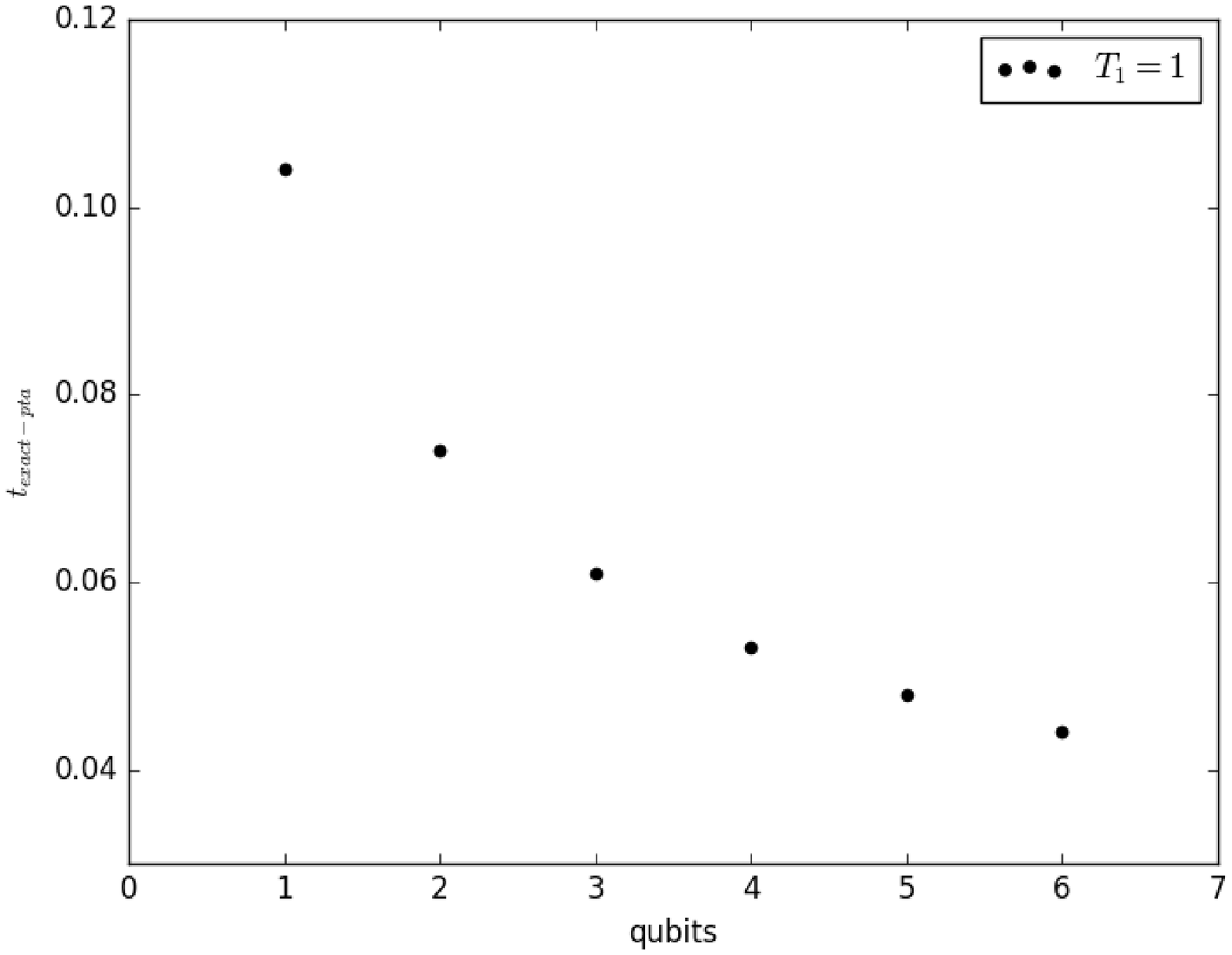}
      \caption{First time at which probabilities differ by more than probabilities of a bit flip}
   \end{figure}
  As can be seen from the second and third figures, the time at which the difference of probabilities surpasses a bit flip probability decreases with the number of qubits. In other words, PTA gets worse earlier in time as qubit number increases.
 \section{Stabilizer Circuits and Quantum Back Action}
  Most of the work done so far on noise models has been done within the context of stabilizer circuits. It is this framework that we assume in this work and argue for how the quantum back-action affects the process quantum of error correction. 
  We can think of a quantum circuit either with Pauli errors or exact noise as one of two CPTPs
  \begin{align}
  &\rho \longmapsto \Lambda_{PTA} \circ G_{n} \circ \Lambda_{PTA}\circ G_{n-1} \dots G_{1} \circ \Lambda_{PTA}(\rho), \\
  &\rho \longmapsto \Lambda_{Exact} \circ G_{n} \circ \Lambda_{Exact}\circ G_{n-1} \dots G_{1} \circ \Lambda_{Exact}(\rho),
  \end{align}
  where \(\Lambda_{Exact} \text{ or } \Lambda_{PTA} \) are the noise models and \(G_{i}\) is the unitary implementation of quantum gates at the \(i^{th}\) step in the quantum circuit.
  By investigating the quantum-back-action we see clearly a relative scale both in time and space.These relative scales in time and space will ultimately depend on what kind of error rates one is willing to tolerate in the computation. In the previous section, the relative scale we chose was one where the probabilities from the two error models first differed by more than a probability of a bit flip.   From the results of the previous section we therefore can conjecture that for small \(n\) i.e small enough circuit depths results from the two error models should not be very different. This is the relative time scale in time. There is also relative time scale in space i.e the dimensional of the Hilbert space that \(G_n \) acts on. This is a slightly less obvious one but none the less should exist.
  
 \subsection{Discrete time quantum error correction and Quantum Back Action}

 \begin{enumerate}
 	\item  For concreteness we assume a stabilizer circuit with distance 3 i.e we can detect and correct one error (this is mainly assumed for simplicity of the discussion).
 	\item There are no state preparation and measurement errors
 \end{enumerate}
 Let us now consider the noise model to be simply Pauli errors for which we allow at most one error in the computation. In this case, there is a one to one map between error syndrome \(S(E)\) and the error \(E\). The main point to understand is that the effect of Quantum Error Correction is to effectively make the measurement process classical. By this we mean the following: in classical measurement theory the point of measurement is merely to find out the state of the system and it is assumed that whatever the answer is, precisely matches what the system was before the measurement. For example if a bit is measured to be in the zero state, it was in the zero state before the measurement. This is precisely what we can't conclude in a quantum measurement, except when are in the context of quantum error correction. That is if for example we have a five qubit code, there are 16 possible error syndrome measurement each corresponding to 16 possible single qubit errors (including no error as a \textit{trivial} error). Should a bit flip occur on the second qubit then the point of measurement is to precisely reflect that. \\
 \indent Now let's assume that we do not have merely Pauli errors but a more physically relevant noise model like amplitude damping. We think of amplitude damping (AD) in the following manner
 $$
 \text{AD = PTA + CROSS TERMS}.
 $$
 If we have a quantum state \(\ket{\Psi} \) then a state with a Pauli error will be \(E\ket{\Psi} = E s_i\ket{\Psi} = s'_i \ket{\Psi} \) where \(s_i \) is a stabilizer of the code and \( s'_i\) is the defective stabilizer. Note that \( [s_j, s'_i] \neq 0 \) and if this was simply PTA we would be done because we would be firmly in the framework of stabilizer formalism. But there are cross terms which have a more realistic model for back action and introduce correlations between Pauli errors.Thus we also lose the property \( \{s_j,s'_i\}=0 \) (anti-commutation) and we are now measuring operators in the stabilizer circuit that neither commute nor anti-commute. The measurement of  \(s'_i\) will affect the dynamics of any other stabilizer \(s_j\). In the same way that measuring the position operator affects measurements of the momentum operator. This why PTA works better for concatenated error correcting codes or when the computation is fault tolerant.\cite{Gutierrez2015,Darmawan2016}. In both these cases the effect is to make the measurement process more classical which is what Quantum Error Correction in stabilizer circuits assumes for the measurement process.
 
 \subsection{Continuous time error correction}
 
 Considering the quantum back action is a more natural question within the context of continuous time error correction where the whole process depends on solving a Lindblad equation. The procedure of error correction is modeled as a quantum jump process. The error correcting procedure infinitesimally is 
 \begin{equation}
 	\rho \rightarrow (1 -\beta dt )\rho + \beta dt \mathcal{R}(\rho),
 \end{equation}
 where \( \mathcal{R} \) is the error correction CPTP map and \( \beta \) is the rate of error correction, while the noise is 
 \begin{equation}
    \rho \rightarrow \rho + \mathcal{L}(\rho) dt + \mathcal{O}(dt^2).
 \end{equation}
Therefore the combined noise and error correction procedure produces the following density matrix equation
\begin{align}
 \frac{d \rho}{dt} = \left( \mathcal{L} + \beta \mathcal{R} -I \right) \rho,
\end{align}
with the formal solution being
\begin{align}
	\rho(t) = e^{(\mathcal{L}+ \beta \mathcal{R} - I)t} \rho(0).
\end{align}
 
 The point is that as we implement \(\mathcal{R} \) through continuous weak measurements we are  assuming a certain error model i.e \( \mathcal{L}_{pta} \) whose Lindblad operators  do not model back-action correctly.  And so rather than seeing a freezing of the state(in the desired quantum state) as we make weak measurements, we should see a reduction in the fidelity as the model for the back-action becomes worse. Considering the back-action could thus be used to produce a time scale for which one can trust the error correction procedure . 
 
   \section{Conclusion}
   Within the context of free qubit dynamics we have argued for general rough features regarding the nature of the Pauli Twirling Approximation by introducing a physical measure namely the probability of no emissions. This measure captures the inherent consequences of the terms dropped by the approximation and can be used to make a prediction as to when they might matter. With this measure we have seen that in the limit of low decoherence and or short times the performance of PTA tracks that of Amplitude Damping but for long times PTA predicts the probabilities that widely diverge from Amplitude Damping. We have also seen that performance of PTA gets worse earlier in the evolution as qubit number increases. It will be the subject of further work to investigate more carefully the relationship between usual measures like fidelity or diamond norms; especially how the back-action measure varies with logical error rates.

   \bibliography{Second_pta_paper}

\end{document}